\PassOptionsToPackage{unicode}{hyperref}
\PassOptionsToPackage{hyphens}{url}
\documentclass[
]{article}
\title{The Transform-o-meter}
\usepackage{etoolbox}
\makeatletter
\providecommand{\subtitle}[1]{
  \apptocmd{\@title}{\par {\large #1 \par}}{}{}
}
\makeatother
\subtitle{A method to forecast the transformative impact of innovation}
\author{Héctor G. T. Torres\footnote{Instituto Tecnológico Autónomo de
  México (ITAM); Email:
  \href{mailto:hgomezt1@itam.mx}{\nolinkurl{hgomezt1@itam.mx}}; Github:
  \url{https://github.com/LornartheBreton}}}
\date{}

\usepackage{amsmath,amssymb}
\usepackage{lmodern}
\usepackage{iftex}
\ifPDFTeX
  \usepackage[T1]{fontenc}
  \usepackage[utf8]{inputenc}
  \usepackage{textcomp} 
\else 
  \usepackage{unicode-math}
  \defaultfontfeatures{Scale=MatchLowercase}
  \defaultfontfeatures[\rmfamily]{Ligatures=TeX,Scale=1}
\fi
\IfFileExists{upquote.sty}{\usepackage{upquote}}{}
\IfFileExists{microtype.sty}{
  \usepackage[]{microtype}
  \UseMicrotypeSet[protrusion]{basicmath} 
}{}
\makeatletter
\@ifundefined{KOMAClassName}{
  \IfFileExists{parskip.sty}{%
    \usepackage{parskip}
  }{
    \setlength{\parindent}{0pt}
    \setlength{\parskip}{6pt plus 2pt minus 1pt}}
}{
  \KOMAoptions{parskip=half}}
\makeatother
\usepackage{xcolor}
\IfFileExists{xurl.sty}{\usepackage{xurl}}{} 
\IfFileExists{bookmark.sty}{\usepackage{bookmark}}{\usepackage{hyperref}}
\hypersetup{
  pdftitle={The Transform-o-meter},
  pdfkeywords={ai, artificial intelligence, transformative artificial
intelligence, general artificial intelligence, innovation},
  hidelinks,
  pdfcreator={LaTeX via pandoc}}
\usepackage{longtable,booktabs,array}
\usepackage{calc} 
\usepackage{etoolbox}
\makeatletter
\patchcmd\longtable{\par}{\if@noskipsec\mbox{}\fi\par}{}{}
\makeatother
\IfFileExists{footnotehyper.sty}{\usepackage{footnotehyper}}{\usepackage{footnote}}
\makesavenoteenv{longtable}
\providecommand{\tightlist}{%
  \setlength{\itemsep}{0pt}\setlength{\parskip}{0pt}}
\newlength{\cslhangindent}
\newlength{\csllabelwidth}
\newlength{\cslentryspacingunit} 
\newenvironment{CSLReferences}[2] 
 {
  \setlength{\parindent}{0pt}
  \ifodd #1
  \let\oldpar\par
  \def\par{\hangindent=\cslhangindent\oldpar}
  \fi
  \setlength{\parskip}{#2\cslentryspacingunit}
 }%
 {}
\usepackage{calc}

\ifLuaTeX
  \usepackage{selnolig}  
\fi

\begin{document}
\maketitle
\begin{abstract}
With the advent of Transformative Artificial Intelligence, it is now
more important than ever to be able to both measure and forecast the
transformative impact/potential of innovation. However, current methods
fall short when faced with this task. This paper introduces the
Transform-o-meter; a methodology that can be used to achieve the
aforementioned goal, and be applied to any innovation, both material and
immaterial. While this method can effectively be used for the mentioned
purpose, it should be taken as a first approach; to be iterated,
researched, and expanded further upon.
\end{abstract}


\thispagestyle{empty}
\clearpage
\pagenumbering{roman}
\tableofcontents
\clearpage
\pagenumbering{arabic}
\setcounter{page}{1}

\hypertarget{introduction}{%
\section{1 - Introduction}\label{introduction}}

The Merriam Webster Dictionary defines ``transformative'' as

\begin{quote}
``{[}To{]} cause or be able to cause lasting change in someone or
something''
(\protect\hyperlink{ref-merrian_webster_dictionary_definition_2022}{Dictionary,
2022})
\end{quote}

Following this definition, it comes natural to describe certain
developments, inventions, ideas, and/or discoveries as transformative.
Wether that be the wheel, calculus, the World Wide Web, or even
Communism; they can all be described as transformative in the sense that
they caused lasting change in humanity.

However, the degree in which they were transformative to humanity
remains to be measured. How does the impact of the wheel compares to
that of calculus? And to that of the American Constitution?

With the advent of Transformative Artificial Intelligence (TAI), the
urgency to answer the aforementioned questions becomes apparent. Several
academics have warned about the sooner-than-expected coming of TAI and
of the life changing effects it would have on humankind
(\protect\hyperlink{ref-bostrom_superintelligence_2014}{Bostrom, 2014};
\protect\hyperlink{ref-gruetzemacher_transformative_2022}{Gruetzemacher
\& Whittlestone, 2022};
\protect\hyperlink{ref-karnofksy_background_2016}{Karnofksy, 2016}).
These questions cannot be left unanswered.

However, the existing methods used to measure/forecast the
transformative impact/potential of innovation are inadequate to answer
the questions at hand. In general, they are too specialized, focusing
their methodologies exclusively on either patents
(\protect\hyperlink{ref-lanjouw_quality_1999}{Lanjouw \& Schankerman,
1999}) and/or academic research
(\protect\hyperlink{ref-greenhalgh_research_2015}{Greenhalgh \& Fahy,
2015}). Their focus is too narrow. For the purposes I describe above, a
new methodology needs to be developed.

Thus, paper introduces the Transform-o-meter; a methodology for both
evaluating forecasting the transformative impact/potential of
innovation.

This paper proceeds as follows. In Section 2, the criteria behind the
Transform-o-meter is explained. In Section 3, Transform-o-meter scores
for some innovations (as well as their reasoning) are given as examples.
Section 4 concludes.

\hypertarget{the-transform-o-meter-methodology}{%
\section{2 - The Transform-o-meter
Methodology}\label{the-transform-o-meter-methodology}}

\hypertarget{defining-what-to-forecast}{%
\subsection{2.1 - Defining What to
Forecast}\label{defining-what-to-forecast}}

The goal of the Transform-o-meter is for it to be able to evaluate the
transformative potential and impact of both material and immaterial
inventions/innovations/ideas. Therefore, for the sake of simplicity,
I'll introduce the concept of an \textbf{Innovation Unit} (or
\textbf{IU}). The Transform-o-meter's criteria's goal is to be
applicable to all IUs. Thus, I shall now formalize the definition of an
IU.

\hypertarget{definition-of-an-innovation-unit-iu}{%
\subsubsection{2.1.1 - Definition of an Innovation Unit
(IU)}\label{definition-of-an-innovation-unit-iu}}

\begin{quote}
An \textbf{Innovation Unit} is a specific, named, artificial invention,
development, discovery, and/or idea.
\end{quote}

\hypertarget{the-criteria-behind-the-transform-o-meter}{%
\subsection{2.2 - The Criteria Behind the
Transform-o-meter}\label{the-criteria-behind-the-transform-o-meter}}

The Transform-o-meter evaluates an IU through six parameters. These
parameters where chosen to be applicable to all IUs, past, present and
future.

The parameters act like a rubric. The IU in question is given an integer
score from 1 to 5 in each of the criterion. This score is then
normalized to an integer scale with a maximum score out of 100.

The criteria are as follows:

\begin{itemize}
\tightlist
\item
  \textbf{Super-seedness Protection}
\item
  \textbf{Magnitude of Economic Impact}
\item
  \textbf{Centralization}
\item
  \textbf{Immediacy of impact}
\item
  \textbf{Uniqueness}
\item
  \textbf{Counter-factual impact}\footnote{Special thanks to
    \href{https://www.christophwinter.net/about}{Christoph Winter} for
    suggesting this parameter.}
\end{itemize}

The following sub-section explains each of the criterion, as well as the
reasoning behind each of the possible scores.

\hypertarget{examining-the-parameters}{%
\subsubsection{2.2.1- Examining the
parameters}\label{examining-the-parameters}}

\hypertarget{super-seedness-protecion}{%
\paragraph{2.2.1.1 - Super-seedness
Protecion}\label{super-seedness-protecion}}

Evaluates if this IU been, in it's purest form, has super-seeded by
another IU for the purpose it was originally intended for; and if other
IUs can be used for the exact same purpose.

\hypertarget{scoring}{%
\subparagraph{2.2.1.1.1 - Scoring}\label{scoring}}

\begin{itemize}
\tightlist
\item
  \textbf{1} - The IU has been completely replaced by other, completely
  different, IU; it is useless.
\item
  \textbf{2} - The IU has been mostly replaced by other IUs that take
  inspiration from the original one.
\item
  \textbf{3} - The IU is used for its original purpose in mostly equal
  conjunction with other, later/contemporary IUs.
\item
  \textbf{4} - The IU is, currently, the most dominant tool used for the
  purpose it was created for, although other IUs exist that do the same
  thing but are not as dominant and/or severely depend on this
  particular IU.
\item
  \textbf{5} - The IU is, currently, the most dominant and efficient
  tool used for the purpose it was originally created for. No other
  known IU can compare.
\end{itemize}

\hypertarget{magnitude-of-economic-impact}{%
\paragraph{2.2.1.2 - Magnitude of Economic
Impact}\label{magnitude-of-economic-impact}}

Evaluates how significant were the changes in humanity's economic
activities as a consequence of the development of the IU.

\hypertarget{scoring-1}{%
\subparagraph{2.2.1.2.1 - Scoring}\label{scoring-1}}

\begin{itemize}
\tightlist
\item
  \textbf{1} - The IU has had minimal economic impact.
\item
  \textbf{2} - The economic impact of the IU is significant, but limited
  to a specific area of expertise/research.
\item
  \textbf{3} - The economic impact of the IU is significant and
  wide-reaching across several areas of expertise.
\item
  \textbf{4} - The IU managed to alter the way at least a generation has
  engaged in economic activities.
\item
  \textbf{5} - The IU fundamentally changed the way humanity engages in
  economic activities.
\end{itemize}

\hypertarget{centralization}{%
\paragraph{2.2.1.3 - Centralization}\label{centralization}}

Measures how centralized was the development of the IU.

\hypertarget{scoring-2}{%
\subparagraph{2.2.1.3.1 - Scoring}\label{scoring-2}}

\begin{itemize}
\tightlist
\item
  \textbf{1} - The IU was created by several civilizations/societies
  over an either unspecified, or centuries-long time period.
\item
  \textbf{2} - The IU was created as a decentralized effort by an entire
  civilization in a period no longer than a century.
\item
  \textbf{3} - The IU was created as an uncoordinated effort of
  different people/groups of people over the span of several decades.
\item
  \textbf{4} - The IU was created as a coordinated effort of different
  people/groups of people over the span of several decades.
\item
  \textbf{5} - The IU was created as a coordinated effort of a singular
  person/group of people over a period no longer than a decade.
\end{itemize}

\hypertarget{immediacy-of-impact3}{%
\paragraph[2.2.1.4 - Immediacy of impact]{\texorpdfstring{2.2.1.4 -
Immediacy of
impact\footnote{It ought to be noted that this criterion was of special
  controversy when discussing the methodology. The objections to it can
  be found in this article's \href{index.md\#4---conclusion}{conclusion}}}{2.2.1.4 - Immediacy of impact}}\label{immediacy-of-impact3}}

Evaluates the time taken for the full-impact of the IU to materialize.

\hypertarget{scoring-3}{%
\subparagraph{2.2.1.4.1 - Scoring}\label{scoring-3}}

\begin{itemize}
\tightlist
\item
  \textbf{1} - The full impact of the IU was not felt until centuries
  after its invention.
\item
  \textbf{2} - The full impact of the IU was not felt until no more than
  a century after its invention.
\item
  \textbf{3} - The full impact of the IU was not felt until no more than
  half a century after its invention.
\item
  \textbf{4} - The full impact of the IU was not felt until no more than
  less than quarter of a century after its invention.
\item
  \textbf{5} - The full impact of the IU was not felt until no more than
  a decade after its invention.
\end{itemize}

\hypertarget{uniqueness}{%
\paragraph{2.2.1.5 - Uniqueness}\label{uniqueness}}

Measures how unique/novel the UI is compared to both prior IUs and
contemporary (at the time) IUs.

\hypertarget{scoring-4}{%
\subparagraph{2.2.1.5.1 - Scoring}\label{scoring-4}}

\begin{itemize}
\tightlist
\item
  \textbf{1} - Not novel at all; similar IUs were developed more than a
  century before this one.
\item
  \textbf{2} - Not very novel; similar IUs were developed less than a
  century before this one.
\item
  \textbf{3} - Contemporarily novel; similar IUs were around the same
  time as this one.
\item
  \textbf{4} - Novel; the IU shares minimal, but noticeable similarity
  to other contemporary IUs.
\item
  \textbf{5} - Top of the line; the IU shares little to no similarity to
  other contemporary and previous IUs.
\end{itemize}

\hypertarget{counter-factual-impact}{%
\paragraph{2.2.1.6 - Counter-factual
impact}\label{counter-factual-impact}}

Measures the likelihood in which the IU could be developed by
contemporaries.

\hypertarget{scoring-5}{%
\subparagraph{2.2.1.6.1 - Scoring}\label{scoring-5}}

\begin{itemize}
\tightlist
\item
  \textbf{1} - Other, independent, unrelated peoples developed virtually
  the same IU at around the same time.
\item
  \textbf{2} - Someone working on the same circle developed virtually
  the same IU at around the same time.
\item
  \textbf{3} - If someone else had the same material resources as the
  innovator, it is very probable that it could've invented it.
\item
  \textbf{4} - If someone else had the same material resources as the
  innovator, it is very unlikely that it could've invented it.
\item
  \textbf{5} - If someone else had the same material resources as the
  innovator, it is impossible that it could've invented it.
\end{itemize}

\hypertarget{transform-o-meter-scores-for-some-ius}{%
\section{3 - Transform-o-meter scores for some
IUs}\label{transform-o-meter-scores-for-some-ius}}

The following subsections score 3 IUs evaluated under the
Transform-o-meter methodology, as well as the reasoning behind the
scores. These are provided as examples on how the methodology can be
applied to any IU rather than as definitive scoring.

\hypertarget{the-wheel}{%
\subsection{3.1 - The Wheel}\label{the-wheel}}

\begin{longtable}[]{@{}
  >{\raggedright\arraybackslash}p{(\columnwidth - 4\tabcolsep) * \real{0.32}}
  >{\raggedright\arraybackslash}p{(\columnwidth - 4\tabcolsep) * \real{0.24}}
  >{\raggedright\arraybackslash}p{(\columnwidth - 4\tabcolsep) * \real{0.44}}@{}}
\toprule
\begin{minipage}[b]{\linewidth}\raggedright
Criteria
\end{minipage} & \begin{minipage}[b]{\linewidth}\raggedright
Score
\end{minipage} & \begin{minipage}[b]{\linewidth}\raggedright
Explanation
\end{minipage} \\
\midrule
\endhead
\textbf{Super-seedness} & 5 & The wheel has not been replaced by any
other IU. \\
\textbf{Economic impact} & 5 & Thanks to the wheel, humanity has
transportation, agriculture, etc., all fundamental for humanity. \\
\textbf{Centralization} & 1 & Several cultures developed the wheel at
different time periods, in different parts of the globe. \\
\textbf{Immediacy of impact} & 1 & Timeline; arguably felt since the
development of agriculture. \\
\textbf{Uniqueness} & 5 & No other IU can be described as similar. \\
\textbf{Counter-factual impact} & 1 & Several, unrelated peoples
developed the same IU at different times. \\
\textbf{Overall} & 60 & \\
\bottomrule
\end{longtable}

\hypertarget{the-world-wide-web}{%
\subsection{3.2 - The World Wide Web}\label{the-world-wide-web}}

\begin{longtable}[]{@{}
  >{\raggedright\arraybackslash}p{(\columnwidth - 4\tabcolsep) * \real{0.32}}
  >{\raggedright\arraybackslash}p{(\columnwidth - 4\tabcolsep) * \real{0.24}}
  >{\raggedright\arraybackslash}p{(\columnwidth - 4\tabcolsep) * \real{0.44}}@{}}
\toprule
\begin{minipage}[b]{\linewidth}\raggedright
Criteria
\end{minipage} & \begin{minipage}[b]{\linewidth}\raggedright
Score
\end{minipage} & \begin{minipage}[b]{\linewidth}\raggedright
Explanation
\end{minipage} \\
\midrule
\endhead
\textbf{Super-seedness} & 5 & It's synonymous to the Internet; the most
dominant protocol. \\
\textbf{Economic impact} & 5 & It has fundamentally changed how humans
produce and communicate. \\
\textbf{Centralization} & 3 & The protocol was developed as an iterative
effort from different parties. \\
\textbf{Immediacy of impact} & 1 & Less than 10 years passed from its
development to the Dot-com Boom. \\
\textbf{Uniqueness} & 3 & Similar communication protocols were developed
around the same time (i.e.~Usenet) \\
\textbf{Counter-factual impact} & 4 & Developed thanks to an iterative
process and U.S. government funding. \\
\textbf{Overall} & 90 & \\
\bottomrule
\end{longtable}

\hypertarget{communism-as-defined-by-marx}{%
\subsection{3.3 - Communism (as defined by
Marx)}\label{communism-as-defined-by-marx}}

\begin{longtable}[]{@{}
  >{\raggedright\arraybackslash}p{(\columnwidth - 4\tabcolsep) * \real{0.32}}
  >{\raggedright\arraybackslash}p{(\columnwidth - 4\tabcolsep) * \real{0.24}}
  >{\raggedright\arraybackslash}p{(\columnwidth - 4\tabcolsep) * \real{0.44}}@{}}
\toprule
\begin{minipage}[b]{\linewidth}\raggedright
Criteria
\end{minipage} & \begin{minipage}[b]{\linewidth}\raggedright
Score
\end{minipage} & \begin{minipage}[b]{\linewidth}\raggedright
Explanation
\end{minipage} \\
\midrule
\endhead
\textbf{Super-seedness} & 2 & There are government systems that take
inspiration from Communism, but no strict Communist ``state'' currently
exists. \\
\textbf{Economic impact} & 4 & Communist states changed how societies
produced during the 20th century. \\
\textbf{Centralization} & 5 & Developed by one man (Marx), with editing
help by Engels. \\
\textbf{Immediacy of impact} & 2 & 74 years passed from the publication
of the Communist Manifesto (1948), to the establishment of the Soviet
Union (1922). \\
\textbf{Uniqueness} & 3 & Marx wasn't the first 18th/19th century
philosopher to reject private property. \\
\textbf{Counter-factual impact} & 3 & Being developed in a book, it is
likely someone else could've developed a very similar system. \\
\textbf{Overall} & 63 & \\
\bottomrule
\end{longtable}

\hypertarget{conclusion}{%
\section{4 - Conclusion}\label{conclusion}}

As shown in the previous sections, the Transform-o-meter's methodology
can be utilized to evaluate any IU. That being said, it is best viewed
as a framework to be further developed, researched, and improved upon.

One of this methodology's main features is also one of its biggest
drawbacks: its serves both to measure the transformative impact of past
IUs, and to forecast their future impact. This dual focus on the \emph{a
priori} and the \emph{a posteriori} led to the inclusion of a
controversial criterion: Immediacy of impact. While it may seem
meaningless to evaluate the transformative impact of an IU by it's
temporal closeness to its invention (a criticism I am in agreement
with), this parameter was included for it's theoretical usefulness in
forecasting; particularly in the context of TAI. An AI-related IU that
received a high score in this criterion (and also scores well overall)
would call for significant and urgent attention, as it's score would
signal it's capacity to be part of (or even be) a TAI.

Furthermore, the scores generated by the Transform-o-meter shouldn't be
static. Rather, they should be dynamically updated as new information
related to each IU arises. Therefore, it'd make sense for the
Transform-o-meter to become an AI.

\hypertarget{the-transform-o-meter-as-an-ai}{%
\subsection{4.1 - The Transform-o-meter as an
AI}\label{the-transform-o-meter-as-an-ai}}

Given the limited scope of this paper, the sample of scores given in
\href{index.md\#3---transform-o-meter-scores-for-some-ius}{Section \#3}
were largely discretionary, and are unlikely to be updated after this
article's publication. Thus, it'd make sense to develop a Machine
Learning model that applies this methodology to new IUs.

For this, I propose to scrape the description of each IU from Wikipedia
(as for the IU to have an entry on it would mean it's relevant enough to
be measured by the Transform-o-meter). Each IU/Description pair would
then be scored, and the data fed to an ML model (most likely an XGBoost
model (\protect\hyperlink{ref-chen_xgboost_2016}{Chen \& Guestrin,
2016}) or an Artificial Neural Network). It's development can be
followed at this Github repo:
\url{https://github.com/LornartheBreton/transform-o-meter}.

\hypertarget{acknowledgements}{%
\subsection{Acknowledgements}\label{acknowledgements}}

I'd like to thank Christoph Winter and my classmates in his Law and AI
course taught at ITAM (which this paper was written for) for their
valuable insights and feedback on this article.

\clearpage

\hypertarget{references}{%
\section*{References}\label{references}}
\addcontentsline{toc}{section}{References}

\hypertarget{refs}{}
\begin{CSLReferences}{1}{0}
\leavevmode\vadjust pre{\hypertarget{ref-bostrom_superintelligence_2014}{}}%
Bostrom, N. (2014). \emph{Superintelligence: {Paths}, {Dangers},
{Strategies}}. Oxford University Press.

\leavevmode\vadjust pre{\hypertarget{ref-chen_xgboost_2016}{}}%
Chen, T., \& Guestrin, C. (2016). {XGBoost}: {A} {Scalable} {Tree}
{Boosting} {System}. \emph{Proceedings of the 22nd {ACM} {SIGKDD}
{International} {Conference} on {Knowledge} {Discovery} and {Data}
{Mining}}, 785--794. \url{https://doi.org/10.1145/2939672.2939785}

\leavevmode\vadjust pre{\hypertarget{ref-merrian_webster_dictionary_definition_2022}{}}%
Dictionary, M. W. (2022). Definition of {Transformative}. In
\emph{Merriam Webster Dictionary}.
\url{https://www.merriam-webster.com/dictionary/transformative}

\leavevmode\vadjust pre{\hypertarget{ref-greenhalgh_research_2015}{}}%
Greenhalgh, T., \& Fahy, N. (2015). Research impact in the
community-based health sciences: An analysis of 162 case studies from
the 2014 {UK} {Research} {Excellence} {Framework}. \emph{BMC Medicine},
\emph{13}(1), 232. \url{https://doi.org/10.1186/s12916-015-0467-4}

\leavevmode\vadjust pre{\hypertarget{ref-gruetzemacher_transformative_2022}{}}%
Gruetzemacher, R., \& Whittlestone, J. (2022). The transformative
potential of artificial intelligence. \emph{Futures}, \emph{135},
102884. \url{https://doi.org/10.1016/j.futures.2021.102884}

\leavevmode\vadjust pre{\hypertarget{ref-karnofksy_background_2016}{}}%
Karnofksy, H. (2016). Some {Background} on {Our} {Views} {Regarding}
{Advanced} {Artificial} {Intelligence}. In \emph{Open Philanthropy}.
\url{https://www.openphilanthropy.org/research/some-background-on-our-views-regarding-advanced-artificial-intelligence/}

\leavevmode\vadjust pre{\hypertarget{ref-lanjouw_quality_1999}{}}%
Lanjouw, J., \& Schankerman, M. (1999). \emph{The {Quality} of {Ideas}:
{Measuring} {Innovation} with {Multiple} {Indicators}} (No. w7345; p.
w7345). National Bureau of Economic Research.
\url{https://doi.org/10.3386/w7345}

\end{CSLReferences}

\end{document}